\documentclass[seceqn,dvips]{arxstspdf}
\usepackage{graphics}
\usepackage{flushend}
\usepackage{stfloats}

\volume{24}
\issue{3}
\pubyear{2009}
\firstpage{255}
\lastpage{269}
\doi{10.1214/09-STS277}

\begin{document}
\begin{frontmatter}

\title{Likelihood Inference for Models with Unobservables: Another View}
\relateddoi{}{Discussed in \doi{10.1214/09-STS277A},
\doi{10.1214/09-STS277B}
and
\doi{10.1214/09-STS277C};
rejoinder at
\doi{10.1214/09-STS277REJ}.}

\begin{aug}
\author[a]{\fnms{Youngjo} \snm{Lee}\ead[label=e1]{youngjo@snu.ac.kr}\corref{}}
\and
\author[b]{\fnms{John A.} \snm{Nelder}\ead[label=e2]{j.nelder@imperial.ac.uk}}

\affiliation{Seoul National University and Imperial College}

\address[a]{Youngjo Lee is a Professor, Department of Statistics,
Seoul National University, Seoul, Korea \printead{e1}.}
\address[b]{John A. Nelder is a Visiting Professor, Department of
Mathematics, Imperial College,
London, SW7 2AZ, UK \printead{e2}.}

\end{aug}

%
\begin{abstract}
There have been controversies among
statisticians on (i) what to model and (ii) how to make inferences from
models with unobservables. One such controversy concerns the difference
between estimation methods for the marginal means not necessarily
having a
probabilistic basis and statistical models having unobservables with a
probabilistic basis. Another concerns likelihood-based inference for
statistical models with unobservables. This needs an extended-likelihood
framework, and we show how one such extension, hierarchical likelihood,
allows this to be done. Modeling of unobservables leads to rich classes of
new probabilistic models from which likelihood-type inferences can be made
naturally with hierarchical likelihood.
\end{abstract}

%
\begin{keyword}
\kwd{Hierarchical generalized linear model}
\kwd{unobservables}
\kwd{random effects}
\kwd{likelihood}
\kwd{extended likelihood}
\kwd{hierarchical likelihood}.
\end{keyword}

\end{frontmatter}

\section{Introduction}

Fisher introduced the
concept of likelihood in 1921 for inferences from statistical models
involving two kinds of objects, namely observed random variables (data) and
unknown fixed parameters. Pearson (\citeyear{1920Pearson}) points out a limitation of Fisher
likelihood for the prediction of unobserved future observations. Fisher's
likelihood cannot be used to make inferences about \mbox{unobservables}. There has
been an effort to extend likelihood inferences to models with unobservables
by eliminating them via integration. However, with a few exceptions
such as
the copula (Joe, \citeyear{1997Joe}), marginal distributions for counts and proportions
are not available in explicit forms, and this restricts the scope of the
classical likelihood approach.

In longitudinal studies, generalized estimating equations (GEEs)
are widely used. They give an estimation method for regression coefficients
constructed directly to describe marginal means with the covariance
structure regarded as contributing nuisance parameters only. However, GEEs
cannot (generally) be integrated to obtain a likelihood function (McCullagh
and Nelder, \citeyear{1989McCullagh}) and therefore may not have a probabilistic or likelihood
basis. These estimation methods for marginal (or population-average) means
are often contrasted with conditional (or subject-specific) models which
include the modeling of unobservables. Jansen~et~al. (\citeyear{2006Jansen})
review the use of GEE methods and conditional models for analysis of
missing data and discuss the choice between them. However, we believe that
such a choice is inappropriate because the choice of an estimation method
for a particular parameterization (marginal parameter) should not pre-empt
the process of model selection. Recently, Lee and Nelder (\citeyear{2004Lee}) have shown
that alleged differences in the behavior of parameters between GEE methods
and conditional models are based on a failure to compare like with
like. We
dislike the use of estimation methods without a probabilistic basis
because, for example, inferences for joint and conditional
probabilities are
not possible.

Recently, broad classes of new probabilistic models with unobserved random
variables (unobservables) have been proposed, such as generalized linear
models (GLMs) with random effects (Lee and Nelder, \citeyear{1996Lee}), latent processes
(Skrondal and Rabe--Hesketh, \citeyear{2009Skrondal}), models for missing data (Little
and Rubin, \citeyear{2002Little}), prediction (Bj\o rnstad, \citeyear{1990Bj}) and for potential outcomes in
causality (Rubin, \citeyear{2006Rubin}), etc. In the statistical literature unobservables
appear with various names such as random effects, latent processes, factor,
missing data, unobserved future observations, potential outcomes, and
so on. Random
effects in the mean model have been proposed to account for within-subject
correlation in longitudinal studies (\citeauthor{1996Diggle}, \citeyear{1996Diggle}) (for
smooth spatial data, see Besag and Higdon, \citeyear{1999Besag}; for spline-type function
fitting, see Eilers and Marx, \citeyear{1996Eilers}; and for factor analysis, see
Bartholomew, \citeyear{1987Bartholomew}, etc.) while random effects in the dispersion model
(Lee and Nelder, \citeyear{2006aLee})
can account for heteroskedasticity, giving heavy-tailed distributions that
allow robust modeling (Noh and Lee, \citeyear{2007aNoh}).

Modeling of unobservables is the key to these new models. However, because
of difficulties in making likelihood inferences about unobservables, some
authors use the Fisher likelihood for inferences about fixed unknown
parameters while for inferences about unobservables they use the empirical
Bayesian (EB) approach or the full Bayesian (FB) inference. Recently,
Zhao et al. (\citeyear{2006Zhao}) have used an FB approach with which they claim to have
an advantage over the frequentist version (EB) in that it is
\textit{computationally simpler} to obtain variance estimates of the random-effect
estimates. (Note that the word ``prediction'' has often been used to denote
the estimation of random effects. However, we believe that it is
clearer to
use \textit{prediction} when we estimate future observations (unobservables)
and \textit{estimation} for the estimation of random effects in the data
already observed.) Discussing the controversy between Fisher and Neyman,
Rubin (\citeyear{2005Rubin}) maintains that models with unobservables arose most naturally
in causal inference within an FB framework. From Lindley and Smith (\citeyear{1972Lindley})
onwards, FB has become dominant for the analysis of models with
unobservables. The availability of Markov-Chain Monte Carlo, which
implements FB procedures, has made FB inferences popular.

By contrast we believe that modeling of unobservables is natural
within an
extended likelihood framework. Recently, for general inferences from models
involving unobservables, Lee and Nelder (\citeyear{1996Lee}) propose the use of the
hierarchical (or h-)likelihood. The h-likelihood plays a key role in the
synthesis of the likelihood inferential tools needed for a broad class of
new models having unobservables. The h-likelihood approach takes into
account the uncertainty in the estimation of random effects, so that
inferences about unobservables are possible without resorting to an EB
framework.

In the next section we review some models with unobservables and discuss
related modeling issues. We review the h-likelihood procedure for the
estimation of random effects and compare them with the Bayesian
approach in
Section~\ref{sec3}; likelihood inferences from such models are demonstrated with
examples in Section~\ref{sec4}, followed by conclusions in Section \ref{sec5}.

\section{How to Model Unobservables}\label{sec2}

Multivariate distributions for non-Gaussian models can be produced by
probabilistic modeling of unobservables without requiring explicit
multivariate generalizations of non-Gaussian distributions. Using
hierarchical likelihood, inferences from these new classes can be
made.

\subsection{HGLMs: Random Effects in the Mean}\label{sec2.1}

HGLMs allow a synthesis of GLMs, random-effect models and
structured-dispersion models. Consider a GLM with random effects where the
response $y$ follows the GLM, conditioning on random effects $v$,
%
\begin{equation}\label{eq:moment}
\mu=\mathrm{E}(y|v)\quad\mbox{and}\quad\operatorname{var}(y|v)=\phi V(\mu)
\end{equation}
with a linear predictor,
%
\begin{equation} \label{eq:hglm}
\eta=X\beta+Zv,
\end{equation}
where $\eta=g(\mu)$ for some monotonic function $g(\cdot)$. When $v$ are normal
the models are called generalized linear mixed models (GLMMs). The use of
other distributions for the random effects enriches the class of
models. Lee
and Nelder (\citeyear{1996Lee}) introduce HGLMs in which the distribution of the random
components is extended to an arbitrary conjugate distribution of a GLM
family with an appropriate link, not necessarily that of the conjugate
pair. Above we suppress the indices to mean that our discussion covers
various models having single or multiple random effects with nested,
crossed, combined structures, etc. We write indices if necessary.

\begin{figure*}[b]

\includegraphics{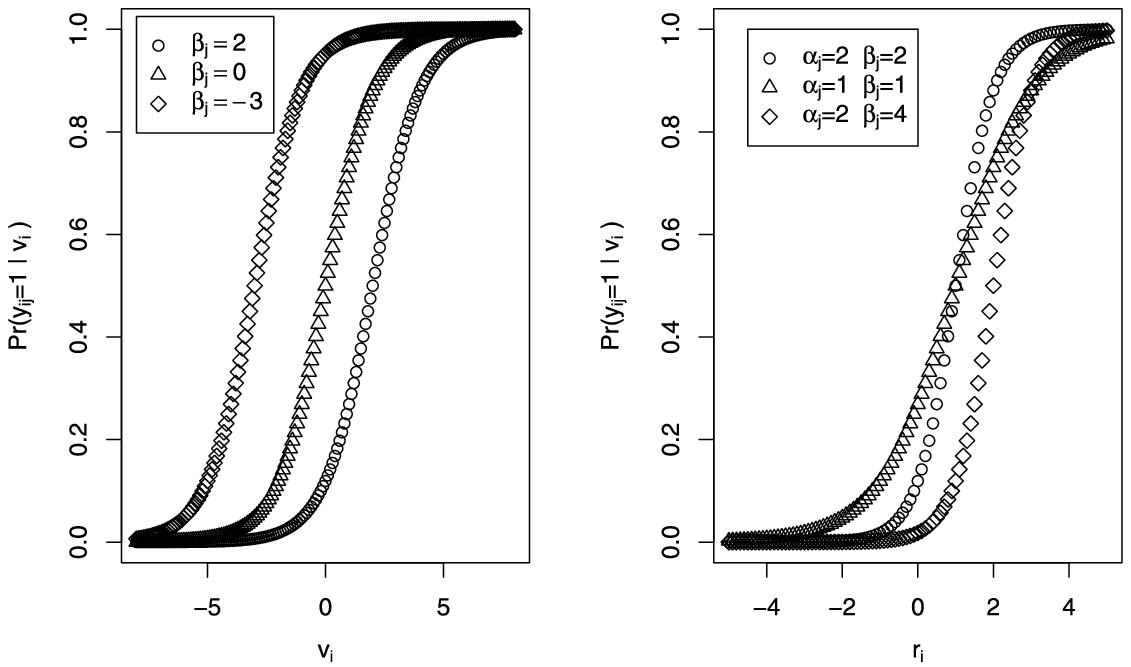}

\caption{Curves of $\operatorname{Pr}(y_{ij}=1|v_{i})$ with respect to $v$
(left) in
a one-parameter IRT model and $r$ (right) in a two-parameter IRT
model.}\label{fig1}
\end{figure*}

To allow various patterned associations among random effects Lee and Nelder
(\citeyear{2001bLee}) propose adding an additional feature to HGLMs as follows: Let $v=Lr$
with $r$ being random effects with a diagonal covariance matrix
$\operatorname{var}(r)=\Lambda$ to give
\begin{eqnarray*}
\operatorname{var}(v)=\Sigma=L\Lambda L^{t}.
\end{eqnarray*}
The last equation can be a spectral decomposition with an orthogonal
matrix $%
L$ or a Choleski decomposition with upper or lower triangular matrix $L.$
Zhao et al. (\citeyear{2006Zhao}) note that the full generality of the
GLMM
requires using general design matrices for both fixed and random components.
With fixed $L,$ not depending upon unknown parameters, we have models for
longitudinal studies, intrinsic autoregressive models, various spline models,
etc. With parameter-dependent $L$ we have random-slope models,
autoregressive models, antedependence models, Markov-random-field
models, and so on (Lee and Nelder, \citeyear{2001bLee}). These models are also able
to handle a great
range of complications in regression-type analysis, for instance,
within-subject correlation in longitudinal data, scatterplot smoothing,
generalized additive models, Kriging, function estimation and non-parametric
regression models such as generalized additive models and
varying-coefficient models (Zhao et al., \citeyear{2006Zhao}).

Example 1: Consider the model from item-response theory (IRT) such that
\begin{eqnarray*}
\mbox{Pr}(y_{ij}=1|v_{ij})=\frac{\exp(v_{ij}-\beta_{j})}{1+\exp
(v_{ij}-\beta_{j})},
\end{eqnarray*}
where $\beta_{j}$ is the intrinsic difficulty of the $j$th item, and $v_{ij}$
is the $i$th subject's ability for the $j$th item. If $v_{ij}=v_{i}$
with $%
v_{i}\thicksim N(0,\lambda)$ it becomes a one-parameter IRT model
(Rasch, \citeyear{1960Rasch}). An appealing feature of this model is that items and subjects
(examinees) can be placed on a common scale. Differences in both difficulty
between items and ability of subjects is assumed to remain the same. In this
model, for a given item, the probability of a correct response increases
monotonically with ability as in Figure \ref{fig1}.

If $v_{ij}=r_{i}\alpha_{j}$ with $r_{i}\thicksim N(0,\lambda)$ and
$\alpha
_{j}$ fixed unknown, we have a two-parameter IRT model. Let $%
v_{i}=(v_{i1},\ldots,v_{ik})^{t}$ and $L_{i}=(\alpha_{1},\ldots
,\alpha
_{k})^{t},$ giving
\begin{eqnarray*}
\operatorname{var}(v_{i})=\Sigma_{i}=L_{i}\Lambda L_{i}^{t},
\end{eqnarray*}
where $\Lambda=\lambda$ is a one-by-one matrix. This model allows for
correlations among items for each subject. In this model $\alpha_{j}$ is
called the discriminant parameter and $\beta_{j}^{\ast}=\beta
_{j}/\alpha
_{j}$ the difficulty parameter (Skrondal and Rabe-Hesketh, \citeyear{2009Skrondal}). This
two-parameter IRT model may lack the monotonicity property in that one item
can be easier than another for one subject, while being more difficult for
another; this is described by the item-subject interaction $r_{i}\alpha
_{j}$. This example shows how a particular modeling of the (singular)
covariance matrix $\Sigma_{i}$ can give an interesting interpretation of
the parameters.


Example 2: When $v_{t}=\rho v_{t-1}+r_{t}$ with $\mbox
{var}(r_{t})=\lambda$
we have autoregressive random effects of order 1. When $\rho=1$ we
have the
random-walk model which gives a singular precision matrix. This random-walk
model for temporal correlation has been extended to spatially-correlated
models via intrinsic autoregressive models with a singular fixed-precision
matrix (Besag and Kooperberg, \citeyear{1995Besag}). Splines can be viewed as
smoothing via
random effects which also have a singular fixed-precision matrix (Green and
Silverman, \citeyear{1994Green}).

Example 3: Skrondal and Rabe-Hesketh (\citeyear{2004Skrondal}) propose generalized linear
latent and mixed models (GLLAMMs) as a means of unifying factor models,
linear structural-relations models and covariate measurement-error models.
They point out that\break GLLAMMs consist of two building blocks, a response model
and a structural model. For the response\break model, they use the HGLM shown in
equation (\ref{eq:hglm}). For the structural model, the random effect itself
satisfies a regression model of the form
\begin{eqnarray*}
v=Bv+\Gamma w+r,
\end{eqnarray*}
where $B$ is a matrix of structural parameters relating the latent dependent
variables to each other, $\Gamma$ is a matrix of structural parameters
relating the latent dependent variable to the latent explanatory variables
and $r$ is a vector of disturbances. From this we have
\begin{eqnarray*}
v=(I-B)^{-1}\Gamma w+(I-B)^{-1}r.
\end{eqnarray*}
Thus, the GLLAMMs can be represented as an\break HGLM with two random components,
\begin{eqnarray*}
\eta=g(\mu)=X\beta+Zv=X\beta+ZL_{1}w+ZL_{2}r,
\end{eqnarray*}
where $L_{1}=(I-B)^{-1}\Gamma$ and $L_{2}=(I-B)^{-1}.$ In\break GLLAMMs the
parametrization using $B$, $\Gamma$, $\operatorname{var}(w)$ and $\operatorname{var}(r)$ gives a useful
interpretation.

Another class of widely used models with unobservables is nonlinear
mixed-effect models in population pharmacokinetics and pharmacodynamics,
models for missing data and models for potential outcomes.

\subsection{Random-Effect Models for the Dispersion}

Lee and Nelder (\citeyear{2006aLee}) introduce double HGLMs (DHGLMs) which allow random
effects for the dispersion. This gives a systematic way of generating
heavy-tailed distributions for various types of data such as counts,
proportions, and so on. Random effects in the mean affect the first two
cumulants
of the distribution of responses while those in the dispersion affect the
third and fourth cumulants, so that by allowing random effects in both mean
and dispersion we can generate models with various patterns in the first
four cumulants. Castillo and Lee (\citeyear{2009Castillo}) show that DHGLMs provide a general
treatment of Levy-process models in financial modeling while Noh and Lee
(\citeyear{2007aNoh}) show that this new class allows robust modeling of GLM classes
with bounded influence. Yun and Lee (\citeyear{2006Yun}) show how to model abrupt
changes in the behavior of schizophrenics. Glidden and Liang (\citeyear{2002Glidden})
show that sensitivity of estimators for $\beta$ from HGLMs become more serious
when the data form a selected sample. However, Noh et al. (\citeyear{2005Noh})
show that by using a heavy-tailed distribution for the random effects,
such a sensitivity in the estimators can be avoided.

\subsection{Probabilistic and Nonprobabilistic Methods}

Without introducing random effects the GEE can be used to obtain maximum
likelihood (ML) estimators when responses are normal. Estimates of
regression coefficients from GEEs have been claimed to be consistent under
various model misspecifications. It is often called the population-averaged
model (Zeger et al., \citeyear{1988Zeger}) or the marginal model (Jansen et
al., \citeyear{2006Jansen}) for a particular parameterization [regression coefficients for
marginal means E($y)$]. For correlated non-normal responses, given a
GEE $%
U(\beta_{s})=\partial q/\partial\beta_{s}=0$ (let us say), the mixed
derivatives may not be the same (McCullagh and Nelder, \citeyear{1989McCullagh}, page 337),
that is,
\begin{eqnarray*}
\partial^{2}q/\partial\beta_{s}\partial\beta_{r}&=&\partial U(\beta
_{s})/\partial\beta_{r}\neq\partial U(\beta_{r})/\partial\beta
_{s}
\\
&=&\partial^{2}q/\partial\beta_{r}\partial\beta_{s};
\end{eqnarray*}
if so there is no probabilistic model leading to the GEE $U(\beta_{s})=0$.
Without such a basis the claim of consistency is meaningless (for more
discussion see Crowder, \citeyear{1995Crowder} and Chaganty and Joe, \citeyear{2006Chaganty}).

It is of interest to study the class of marginal models, allowing
estimating equations. Various marginal models have been proposed by
Molenberghs and\break Lesaffre (\citeyear{1994Molenberghs}), Molenberghs et al. (\citeyear{2007Molenberghs}) and
Heagerty and Zeger (\citeyear{2000Heagerty}). Heagerty and Zeger (\citeyear{2000Heagerty}) claimed that the
parameter estimates from their\break marginal models were less sensitive to the
misspecification of the distribution of random effects. Lee and Nelder
(\citeyear{2004Lee}) show that if one compares like with like the differences between
the results from the two models are not great. All that we can say is that
certain parameterizations are less sensitive under certain probabilistic
models so that it could be recommended to use such a parameterization
if it
also met scientific requirements. For further controversies on
parameterizations see Lindsey and Lambert (\citeyear{1998Lindsey}).

GEE is an estimating \textit{method}, not a model. Thus we do not believe
that a useful comparison can be made between a probabilistic model such
as a
HGLM and an estimating method such as GEE. We see the analysis of data as
consisting of three main activities: the first two are model fitting and
model checking which aim to find parsimonious well-fitting models, and
together comprise model selection; the third is model prediction, where
parameter estimates from selected models are used to predict quantities of
interest and their uncertainties. In our view, inferences about margins and
individual subjects' responses and a choice of an estimation method
such as
the GEE, ML, etc., both belong to the prediction phase of the analysis.

In this paper we shall not consider GEE further because the method does not
allow inferences about unobservables.

\section{Extended Likelihood versus Bayesian Approaches}\label{sec3}

Besides the observed data and fixed unknown parameters in Fisher likelihood,
an additional type of object, namely unobservable random variables $v,$ is
often of interest in making statistical inferences.

Example 4: Suppose that we have the number of epileptic seizures in an
individual for five weeks, $y=(3,2,5,0,4)$. Suppose that these counts are
i.i.d. from a Poisson distribution with mean $\theta.$ Now we want to have
a predictive probability function for the seizure counts for the next
week $v
$. Here, $\hat{\theta}$ $=(3+2+5+0+4)/5=2.8,$ so that the plug-in technique
gives the predictive distribution for the seizure count $v$ of the next
week
\begin{eqnarray*}
f_{\hat{\theta}}(v=i|y)=f_{\hat{\theta}}(v=i)=\exp(-2.8)2.8^{i}/i!.
\end{eqnarray*}
Pearson (\citeyear{1920Pearson}) points out the limitation of Fisher likelihood using the
plug-in method because it cannot account for uncertainty in estimating $
\theta.$

Example 5: Suppose that the data $Y$ are collected from the statistical
model $f_{\theta}(Y;\theta).$ Suppose also that some of the intended
observations in $Y$ are unobservable because they are missing. We write
$%
Y=(y_{\mathrm{obs}},y_{\mathrm{mis}})$ for $y_{\mathrm{obs}}$ the observed and $y_{\mathrm{mis}}$ for the missing
components. Let $r$ be missing data indicators such that
\begin{eqnarray*}
r_{i} =\cases{ 1,&\mbox{if }$Y_{i}$\mbox{ is missing,}
\cr
0,&\mbox{if }$Y_{i}$\mbox{ is observed.}
}
\end{eqnarray*}
This leads to a probability function
\begin{eqnarray*}
f_{\theta}(Y,r;\theta)\equiv f_{\theta}(Y)f_{\theta}(r|Y).
\end{eqnarray*}
Here $y=(y_{\mathrm{obs}},r)$ are the observed data, and $y_{\mathrm{mis}}$ are the
unobservables.

From these models, likelihood inferences can be made using the h-likelihood
defined by
%
\begin{eqnarray} \label{eq:h-likelihood}
h&=&h(\theta,v)=\log f_{\theta}(y|v)+\log f_{\theta}(v)\nonumber
\\[-8pt]\\[-8pt]
&=&\log f_{\theta
}(y,v)=m+\log f_{\theta}(v|y)\nonumber
\end{eqnarray}
where $m$ is the marginal log-likelihood $m=\log f_{\theta}(y)$ with $%
f_{\theta}(y)=\int f_{\theta}(y|v)f_{\theta}(v)\,dv.$ This is the (log)
h-likelihood which plays the same role as the\break log-likelihood $m$ in
Fisher's likelihood inference for models without unobservables. In forming
the h-likelihood the choice of the scale for $v$ is important (Lee et
al., \citeyear{2006Lee}) because the mode and its curvature are used for inferences
as we shall discuss.

Throughout this paper we use $f_{\theta}(\cdot)$ to denote probability functions
of random variables with fixed parameters $\theta$; the arguments within
the brackets can be either conditional or unconditional. Thus $f_{\theta
}(y|v)$ and $f_{\theta}(v|y)$ have different functional forms though
we use
the same $f_{\theta}(\cdot)$ to mean probability functions with parameters $
\theta$.

\subsection{Bayesian Inferences}

\noindent If we assume a prior $\pi(\theta)$ on parameters $\theta$,
Bayesian inferences can be made. The posterior is
\begin{eqnarray*}
\pi(\theta,v|y)\varpropto\pi(y|v,\theta)\pi(v|\theta)\pi(\theta),
\end{eqnarray*}
where $\pi(y|v,\theta)=f_{\theta}(y|v)$ and $\pi(v|\theta
)=f_{\theta
}(v).$ Here $\theta$ is also unobservable and is eliminated by integration.
Let $\theta_{-i}=(\theta_{1},\ldots,\theta_{i-1},\theta
_{i+1},\ldots
,\theta_{p})^{T}.$ For\break Bayesian inferences the following various
marginal or conditional
posteriors have been used:\
\begin{eqnarray*}
\pi(\theta|y)&=&\int\pi(\theta,v|y)\,dv,
\\
\pi(\theta_{i}|y)&=&\int\pi(\theta,v|y)\,dv\,d\theta_{-i},
\\
\pi(v_{i}|y)&=&\int\pi(\theta,v|y)\,dv_{-i}\,d\theta,
\\
\pi(v_{i}|y,\theta)&=&\int\pi(v|y,\theta)\,dv_{-i}.
\end{eqnarray*}
 In this
paper full Bayesian (FB) inference is assumed to use the marginal posteriors
$\pi(\theta_{i}|y)$ and $\pi(v_{i}|y)$ while empirical Bayesian (EB)
inference (Morris, \citeyear{1983Morris}) uses the conditional posteriors $\pi
(v_{i}|y,\hat{%
\theta})$ where $\hat{\theta}$ are the ML estimators maximizing the
likelihood $f_{\theta}(y)=\pi(\theta|y)=\int\pi(\theta,v|y)\,dv$ under
the uniform prior $\pi(\theta)=1.$

\subsection{Adjusted Profile H-likelihoods and Likelihood Inference}

The likelihood principle of Birnbaum (\citeyear{1962Birnbaum}) states that Fisher's marginal
likelihood $f_{\theta}(y)$ carries all the (relevant experimental)
information in the data about the fixed parameters $\theta$ so that $%
f_{\theta}(y)$ should be used for inferences about $\theta$ (see also
Berger and Wolpert, \citeyear{1984Berger}). For estimating fixed parameters $\theta$ we
follow the likelihood principle by using the ML estimator from
$f_{\theta
}(y).$ We view the marginal likelihood as an adjusted profile likelihood
eliminating nuisance unobservables $v$ from the h-likelihood. However, the
computation of ML estimators can be a complex task because of intractable
integration. For example, in the Salamander data (McCullagh and Nelder,
\citeyear{1989McCullagh}) marginal-likelihood inference, based upon numerical integration using
Gauss--Hermite quadrature, is not feasible since a\break $120$-dimensional integral
is required.

Let
\begin{eqnarray*}
\ell=\ell(\theta)=\log f_{\theta}(y)=\log\int\exp h\, dv
\end{eqnarray*}
be the (log-) marginal likelihood. Let $l=l(\alpha,\psi)$ be a likelihood,
either a marginal likelihood $\ell$ or an hierarchical likelihood $h,$ with
nuisance parameters $\alpha.$ Lee and Nelder (\citeyear{2001aLee}) introduce a
function, $%
p_{\alpha}(l;\psi)$, defined by
%
\begin{eqnarray}
&&p_{\alpha}(l;\psi)\nonumber
\\[-8pt]\\[-8pt]
&&\quad =\biggl[l-\frac{1}{2}\log\det\{D(l,\alpha)/(2\pi
)\}\biggr]\bigg|_{\alpha=\tilde{\alpha}},\nonumber
\end{eqnarray}
where $D(l,\alpha)=-\partial^{2}l/\partial\alpha^{2}$ and $\tilde
{\alpha}
$ solves $\partial l/\partial\alpha=0$. These $p(\cdot)$ functions define
adjusted profile h-likelihoods (APHLs). If $\pi(\theta)=1$ the Bayesian
posterior is identical to the h-likelihood, $\pi(\theta,v|y)=f_{\theta
}(y,v).$ Thus APHLs can have a Bayesian interpretation; for example $%
p_{v_{-i},\theta}(h;v_{i})$ is the Laplace approximation to the marginal
posterior $\pi(v_{i}|y),$ eliminating ($v_{-i},\theta)$ by integration$.$
When $\pi(\theta)=1$, it is not a probability if the domain is the whole
real line or the positive real line. However, as long as the marginal
posterior is proper (finite), $\pi(v_{i}|y)$ would be considered as a valid
posterior (Berger, \citeyear{1985Berger}).

APHLs also allow a likelihood interpretation. Here $p_{v}(h;\theta)$
is the
Laplace approximation to the marginal likelihood $\ell$ obtained by
integrating over unobservables $v$ (Lee and Nelder, \citeyear{2001aLee}); its maximum
gives approximate (marginal) ML estimators for $\beta.$ In likelihood
inferences fixed parameters are eliminated by conditioning (if
available) or
profiling (in general). Now suppose that the parameters in a model can be
divided into location parameters $\beta$ and dispersion parameters
$\sigma
^{2}.$ Note that $p_{\beta}(\ell;\sigma^{2})$ is an adjusted profile
likelihood that approximates the conditional log-likelihood obtained by
conditioning on the marginal ML estimator $\tilde{\beta}$ to eliminate the
fixed unknown parameter $\beta$ (Cox and Reid, \citeyear{1987Cox}). A well-known exact
example of this is the use of restricted likelihood in linear mixed models.
Furthermore,\break $p_{\theta}(h;v)$ is Davison's (\citeyear{1986Davison}) predictive likelihood
for $v$, eliminating nuisance fixed parameters $\theta$. The APHL $%
p_{v_{-i},\theta}(h;v_{i})$ eliminates $v_{-i}$ by integration and
$\theta$
by conditioning on $\hat{\theta}.$ When orthogonality does not hold between
parameters we use a profile likelihood to eliminate nuisance
parameters. To
simplify the notation we sometimes suppress arguments; for example we
use $%
p_{v}(h)$ instead of $p_{v}\{h(v,\beta,\sigma^{2});\beta,\sigma
^{2}\}=p_{v}(h;\beta,\sigma^{2})$ if this does not lead to
ambiguity.

Lee and Nelder (\citeyear{1996Lee}, \citeyear{2001aLee}, \citeyear{2006aLee}) propose maximizing the h-likelihood $h$
for the estimation of $v$, the marginal likelihood $\ell$ for the ML
estimators for $\beta$ and the restricted likelihood $p_{\beta}(\ell)$
for the dispersion parameters $\sigma^{2}$. Thus our position is
consistent with the likelihood principle by using the marginal likelihood
for inferences about $\theta$. However, when $\ell$ is numerically hard
to obtain, we propose to use adjusted profile h-likelihoods (APHLs) $%
p_{v}(h) $ and $p_{\beta,v}(h)$ as approximations to $\ell$ and
$p_{\beta
}(\ell);$ $p_{\beta,v}(h)$ approximates the restricted log-likelihood.
Second-order Laplace approximations may sometimes be useful to improve
accuracy.

Many numerical studies on h-likelihood have shown that this development
gives practically satisfactory estimates of parameters in many models where
the ML estimators are hard to compute. For binary data Noh and Lee (\citeyear{2007bNoh})
show numerically that the h-likelihood estimator for $\theta$ has less
bias and mean square error than various other methods developed by Schall
(\citeyear{1991Schall}), Breslow and Clayton (\citeyear{1993Breslow}), Drum and McCullagh (\citeyear{1993Drum}), Shun and
McCullagh (\citeyear{1995Shun}), Lin and Breslow (\citeyear{1996Lin}) and Shun (\citeyear{1997Shun}): see also the
simulation studies of frailty models (Ha and Lee, \citeyear{2005Ha}) and of mixed linear
models with censoring (Ha et al., \citeyear{2002Ha}). In the salamander data,
among other methods considered, the MCEM of Vaida and Meng (\citeyear{2004Vaida}) gives the
closest estimates to the h-likelihood estimators.

Little and Rubin (\citeyear{2002Little}) provide an extensive review of the analysis of
missing data and claim that h-likelihood methods are inappropriate for
the estimation of $\theta$ in missing-value settings such as that in
Example~5. They appear to wrongly equate h-likelihood estimation to a
joint maximization of mean and dispersion parameters. Yun et al.
(\citeyear{2007Yun}) show, in contrast to this assertion, that when applied
appropriately h-likelihood methods are both valid and efficient in such
settings. In non-linear mixed-effect models the h-likelihood can also
improve on existing methods (Noh and Lee, \citeyear{2008Noh}).

\begin{figure*}

\includegraphics{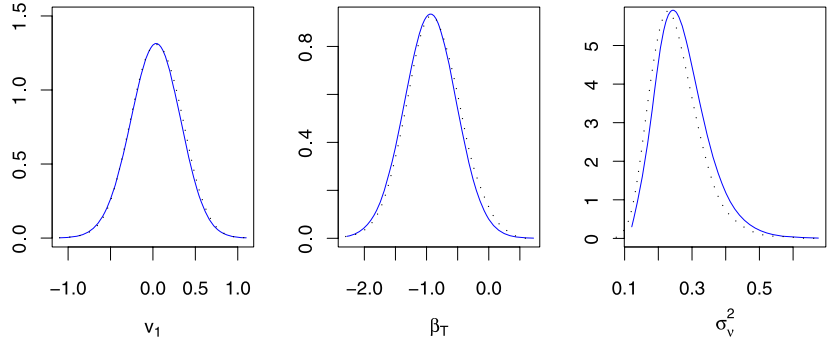}

\caption{The marginal posteriors ($\cdots$) versus APHLs
($-$).}\label{fig2}
\end{figure*}

\subsection{APHLs versus Marginal Posteriors}\label{sec3.3}

In the Bayesian approach, simulation techniques such as MCMC are often used
to compute the\break marginal posteriors. Consider the Epil example of the
OpenBUGS manual, volume 1 (Thomas et al., \citeyear{2006Thomas}). The data come from
a clinical trial of 59 epileptic patients. Each patient $i$ is
randomized to
a new drug ($T_{i}=1)$ or a placebo ($T_{i}=0$). The observations for each
patient $y_{i1},\ldots,y_{i4}$ are the number of seizures during the 2
weeks before each of four visits. The covariates are age ($A_{i}$), the
baseline seizure counts ($B_{i}$) and an indicator variable for the fourth
clinic visit ($V4$). Consider the HGLM,
\begin{eqnarray*}
\eta_{ij}&=&\beta_{0}+\beta_{B}\log(B_{i}/4)+\beta_{T}T_{i}
\\
&&{}+\beta
_{T\times B}T_{i}\times\log(B_{i}/4)
+\beta_{A}A_{i}
\\
&&{}+\beta
_{V}V4
+v_{i}+w_{ij,}
\end{eqnarray*}
using centered covariates with $v_{i}\backsim N(0,\sigma_{v}^{2})$
and\break
$%
w_{ij}\backsim N(0,\sigma_{w}^{2}).$ In discussing the paper by Rue et
al. (\citeyear{2009Rue}) on Bayesian inferences based on priors $\sigma
_{v}^{-2},\sigma_{w}^{-2}\backsim \operatorname{gamma} (0.001,0.001),$ Lee shows Figure
\ref{fig2} (of this paper) for the marginal posteriors, $\pi(v_{1}|y),$ $\pi
(\beta
_{T}|y)$ and $\pi(\sigma_{v}^{2}|y),$ from OpenBUGS (Thomas et
al., \citeyear{2006Thomas}) and the corresponding APHLs, $p_{v_{-1},w,\theta
}(h;\break v_{1}),$ $%
p_{v,w}(h;\beta_{T},\hat{\theta}(\beta_{T}))$ and $p_{v,w,\beta
}(h;\sigma
_{v}^{2}, \hat{\sigma}_{w}^{2}(\sigma_{v}^{2}))$\break where $\hat{\theta
}(\alpha
)$ are the ML estimators of remaining $\beta$ and the REML estimators for
the dispersion parameters at $\beta_{T}=\alpha$ and $\hat{\sigma}%
_{w}^{2}(\alpha)$ is the REML estimators of $\sigma_{w}^{2}$ at
$\sigma
_{v}^{2}=\alpha.$ Figure~\ref{fig2} shows almost identical plots for both
random and
fixed effects. However, the plots for the dispersion components are
different because the inverse-gamma prior of Rue et al. (\citeyear{2009Rue}) is
informative. This leads to biases when dispersion parameters are not random
but are fixed unknowns, as in disease mappings (Jang et al., \citeyear{2007Jang}).
Thus without MCMC samplings similar information could be obtained from the
extended likelihood unless the assumed prior is informative. Thus,
likelihood inferences can be made without the necessity of inventing priors
for parameters.


\section{Likelihood Inference for Unobservables}\label{sec4}

The extended likelihood principle of Bj\o rnstad\break (\citeyear{1996Bj}) shows that extended
likelihood, of which h-likelihood is a special case, carries all the
information in the data about the unobserved quantities $v$ and $\theta.$
Bedrick and Hill (\citeyear{1999Bedrick}) study the use of extended likelihood as a summary
function for\break unobservables. In this paper we discuss its use as an
estimating tool.

\begin{figure*}

\includegraphics{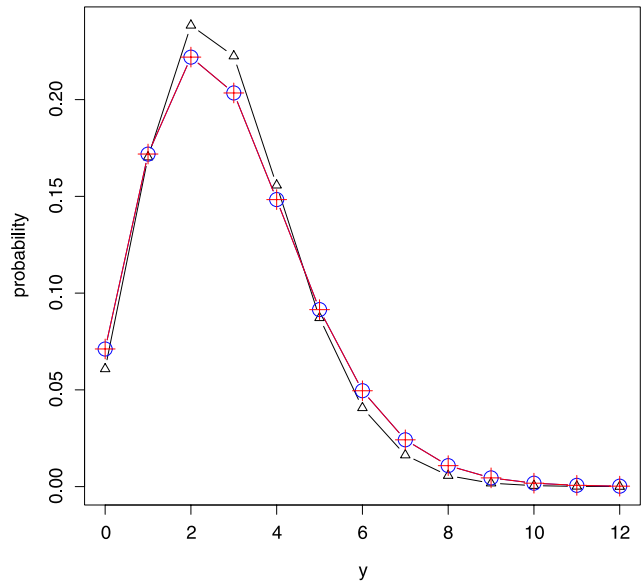}

\caption{Predictive density of the number of seizure counts: Plug-in
method ($\triangle$), Bayesian method ($\circ$) and h-likelihood
method ($+$).}\label{fig3}
\end{figure*}

Consider the prediction problem in Example 4 where the plug-in
technique $%
f_{\hat{\theta}}(v=i)=f_{\hat{\theta}}(v=i|y)=\pi(v=i|y,\hat{\theta})$ can
be viewed as the EB. With Jeffreys' prior, $\pi(\theta)\varpropto$ $%
\theta^{-1/2},$ the resulting marginal posterior $\pi(v|y)$ gives a
predictive probability with higher probabilities for larger $y$. Pawitan
(\citeyear{2001Pawitan}) considers the h-likelihood, proportional to
\begin{eqnarray*}
f_{\theta}(3,2,5,0,4,v)&=&\exp(-6\theta)\theta
^{3+2+5+0+4+v}
\\
&&{}/(3!2!5!0!4!v!).
\end{eqnarray*}
Here $\hat{\theta}(v)=(3+2+5+0+4+v)/6.$ Then the normalized profile
likelihood $f_{\hat{\theta}(v)}(3,2,5,0,4,v)$ gives the predictive
distribution of Mathiasen (\citeyear{1979Mathiasen}) almost identical to Pearson's but without
assuming a prior on $\theta$ (Figure \ref{fig3}) (for more discussion, see Bj\o
rnstad, \citeyear{1990Bj}). This example shows that standard methods for likelihood
inferences can be used for the prediction problem. In the next section we
illustrate how to use standard likelihood methods to overcome a
drawback of
EB method.


\subsection{EB Versus H-likelihood Methods}

Because the Fisher likelihood $f_{\theta}(y)$ in (\ref{eq:h-likelihood})
does not involve $v,$ the other component (the conditional posterior) $%
f_{\theta}(v|y)=\pi(v|y,\theta)$ seems to carry all the information in
the data about the unobservables. Thus an inference would be based solely
upon the estimated posterior,
\begin{eqnarray*}
f_{\hat{\theta}}(v|y)=\pi(v|y,\hat{\theta}),
\end{eqnarray*}
where $\hat{\theta}$ are usually the ML estimators (Carlin and Louis, \citeyear{2000Carlin}).
Using $f_{\hat{\theta}}(v|y)$ to make inferences about $v$ is naive,
and Bj\o
rnstad (\citeyear{1990Bj}) shows how badly it performs in measuring the true
uncertainty in estimating $v$. Note that maximization of the
h-likelihood (%
\ref{eq:h-likelihood}) yields EB-mode estimators for $v$ without
computing $%
f_{\theta}(v|y)$. However, the Hessian matrix based upon the estimated
posterior $f_{\hat{\theta}}(v|y)$ gives a naive variance estimate for the
prediction $\hat{v}$ because it does not properly account for the
uncertainty caused by estimating $\theta$. Note that the marginal posterior
variance is
%
\begin{eqnarray}\label{eq:bayes}
\operatorname{var}(v_{i}|y)&=&E_{\theta|y}[\operatorname{var}(v_{i}|y,\theta)]\nonumber
\\[-8pt]\\[-8pt]
&&{}+\operatorname{var}%
_{\theta|y}[E(v_{i}|y,\theta)].\nonumber
\end{eqnarray}
Carlin and Gelfand (\citeyear{1990Carlin}) note that the naive EB variance estimate only
approximates the first term in the equation above. Laird and Louis (\citeyear{1987Laird})
and Carlin and Gelfand (\citeyear{1990Carlin}) propose to use the bootstrap method to estimate
the second term. In this paper the FB method uses the marginal
posterior $%
\pi(v_{i}|y).$

Up to now most studies on h-likelihood methods have been about the
efficiencies of parameter estimates. Here we discuss how to compute the
variance of estimated random effects. We see that inferences about random
effects cannot be made by using only $f_{\theta}(v|y)$ as the EB method
does. Because $f_{\theta}(v|y)$ involves the fixed parameters $\theta
$ we
should use the whole h-likelihood to reflect the uncertainty in
estimating $%
\theta;$ it is the other component $f_{\theta}(y)$ which carries the
information about this. By using the h-likelihood, complete likelihood
inferences can be made not only for $\theta$ but also for $v$ and their
combinations.

Given $\theta$ let $\hat{v}(\theta)$ be a random-effect estimator
solving $%
\partial h/\partial v=0.$ As a variance of random-effect estimators Booth
and Hobert (\citeyear{1998Booth}) recommend using the conditional mean square error (CMSE)
defined by
%
\begin{eqnarray}\label{eq:cmsep}
\mathit{CMSE}(v)&=&\mathrm{E}\bigl\{\bigl(\hat{v}(\hat{\theta})-v\bigr)\bigl(\hat{v}(\hat{\theta
})-v\bigr)^{\prime
}|y\bigr\}\nonumber
\\[-8pt]\\[-8pt]
&=&\operatorname{var}_{\theta}(v|y)+D(\theta),\nonumber
\end{eqnarray}
where $\operatorname{var}_{\theta}(v|y)=\mbox{E}\{(\hat{v}(\theta)-v)(\hat{v}%
(\theta)-v)^{\prime}|y\}$\break and $D(\theta)=\mbox{E}\{(\hat{v}(\hat
{\theta})-%
\hat{v}(\theta))(\hat{v}(\hat{\theta})-\hat{v}(\theta))^{\prime}|y\}
$ is
the inflation of the CMSE caused by estimating $\theta.$ The EB estimator,
the inverse of the Hessian matrix from $\log f_{\theta}(v|y),$ gives an
estimator for the first term\break $\operatorname{var}_{\theta}(v|y)$ in (\ref
{eq:cmsep})$%
.$ Thus it could give severe underestimation if $D(\theta)$ is large.
Lee and Nelder (\citeyear{1996Lee}) note that in HGLMs (\ref{eq:hglm}), the location
parameters ($v,\beta$) and dispersion parameters $\sigma^{2}=$($\phi
,\Sigma$) are orthogonal so that we need consider only the variance
inflation caused by estimating $\beta.$ The Hessian matrix of $\beta$
and $%
v$ is given by
%
\begin{equation}\label{eq:hessian}
\quad I(\beta,v)=-\left(
\matrix{
\partial^{2}h/\partial\beta\partial\beta^{\prime} & \partial
^{2}h/\partial\beta\partial v^{\prime}
\cr
\partial^{2}h/\partial v\partial\beta^{\prime} & \partial
^{2}h/\partial
v\partial v^{\prime}
}
\right).
\end{equation}
Here the EB variance estimator is given by $-(\partial^{2}h/\break \partial
v\partial v^{\prime})^{-1}|_{\theta=\hat{\theta}}$. Lee and Ha (\citeyear{2008Lee})
show that in general the inverse of the Hessian matrix (\ref{eq:hessian})
gives an approximation to the CMSE (\ref{eq:cmsep}). Before we discuss the
general use of this method we investigate a simple example which shows
issues related to this problem.

\subsection{Bayarri's Example}

Bayarri et al. (\citeyear{1988Bayarri}) try to show by an example that likelihood
inference is not possible for general models with unobservables. Suppose
that there is a single fixed parameter $\theta$, a single unobservable
random quantity $u$ and a single observable quantity $y.$ An unobserved
random variable $u$ has a probability function
\begin{eqnarray*}
f_{\theta}(u)=\theta\exp(-\theta u)\quad\mbox{for }u>0, \theta>0,
\end{eqnarray*}
and an observable random variable $y$ has conditional probability function
\begin{eqnarray*}
f_{\theta}(y|u)=f(y|u)=u\exp(-uy)\quad \mbox{for } y>0, u>0,
\end{eqnarray*}
free of $\theta$. Besides $f(y|u),$ they considered the following two
additional
possibilities for an extended likelihood for models with these three kinds
of objects:
\begin{eqnarray*}
f_{\theta}(y) &=&\frac{\theta}{(\theta+y)^{2}},
\\
f_{\theta}(y,u) &=&u\theta\exp\{-u(\theta+y)\}.
\end{eqnarray*}
The marginal log-likelihood $m=\log f_{\theta}(y)$ gives the ML estimator
for $\theta$ but is totally uninformative about the unknown value of $u$.
The conditional likelihood $f(y|u)$ is uninformative about $\theta$ and
loses the relationship between $u$ and $\theta$ reflected in $f_{\theta
}(u).$ Finally, the extended likelihood $f_{\theta}(y,u)$ yields, if
maximized jointly with respect to $\theta$ and $u$, the useless
estimators $%
\hat{\theta}=\infty$ and $\hat{u}=0$. Bayarri et al. (\citeyear{1988Bayarri})
therefore conclude that none is useful as a likelihood for complete
inferences, so that Bayes is the only method for inferences from general
models.

The h-(log)-likelihood is given by
\begin{eqnarray*}
h&=&\log f_{\theta}(y,v)=\log f_{\theta}(y,u)+\log|du/dv|
\\
&\equiv& 2v+\log
\theta-u(\theta+y),
\end{eqnarray*}
where $v=\log u$ with $v$ being the canonical scale in which the joint
maximization of $h$ with respect to $\theta$ and $u$ gives the ML estimator
of $\theta$ (Lee et al., \citeyear{2006aLee}). Suppose that the marginal
likelihood is hard to obtain. The Laplace approximation is proportional
to $%
m=\log f_{\theta}(y)$ and gives the ML estimator $\hat{\theta}=y$ and its
variance estimator
\begin{eqnarray*}
\widehat{\operatorname{var}(\hat{\theta})}=-\{\partial^{2}m/\partial\theta
^{2}|_{\theta=\hat{\theta}}\}^{-1}=2y^{2}.
\end{eqnarray*}
Given $\theta,$ the estimating equation $\partial h/\partial u=0$
gives the
best estimator of $u$ (Robinson, \citeyear{1991Robinson}),
\begin{eqnarray*}
\hat{u}(\theta)=\mbox{E}(u|y)=\frac{2}{\theta+y},
\end{eqnarray*}
from which we have
\begin{eqnarray*}
\hat{u}(\hat{\theta})=\frac{2}{\hat{\theta}+y}=\frac{1}{y}.
\end{eqnarray*}
Furthermore, we have
\begin{eqnarray*}
I(\theta,\hat{u}(\theta))&=&-\left(
\matrix{
\partial^{2}h/\partial\theta^{2} & \partial^{2}h/\partial\theta
\partial u \cr
\partial^{2}h/\partial u\partial\theta& \partial^{2}h/\partial u^{2}
}
\right)
\\
&=&\left(
\matrix{
1/\theta^{2} & 1 \cr
1 & (y+\theta)^{2}/2
}
\right) .
\end{eqnarray*}
Note here that
\begin{eqnarray*}
\operatorname{var}_{\theta}(u|y)=\mathrm{E}\bigl\{\bigl(\hat{u}(\theta)-u\bigr)^{2}|y\bigr\}
=2/(y+\theta
)^{2}
\end{eqnarray*}
so that EB gives $\widehat{\operatorname{var}_{\theta}(u|y)}=1/(2y^{2}).$ Here
$%
D(\theta)=\mbox{E}[\{1/y-2/(\theta+y)\}^{2}|y\}=(y-\theta
)^{2}/\{y(y+\theta)\}^{2}=(\hat{\theta}-\theta)^{2}/\{y(y+\theta)\}^{2},$
so that, following Booth and Hobert (\citeyear{1998Booth}), if we estimate $(\hat{\theta
}%
-\theta)^{2}$ by var($\hat{\theta})$ we have $\widehat{D(\theta)}%
=2y^{2}/4y^{4}=1/(2y^{2}).$ Thus the estimator for the CMSE is $1/y^{2},$
which can be obtained from the corresponding element in the Hessian
matrix $%
I(\hat{\theta},\hat{u}(\hat{\theta})).$ An alternative justification is that
the h-likelihood variance estimator is estimating the unconditional
mean-square error because\break $\widehat{\mbox{E}\{(\hat{u}(\hat{\theta
})-u)^{2}\}%
}=1/y^{2}$ from $\mbox{E}\{(\hat{u}(\hat{\theta})-u)^{2}\}=1/\theta^{2}$
(Lee et al., \citeyear{2006Lee}, page 116).

With this small example we illustrate how the h-likelihood gives complete
likelihood inferences, giving the ML inference for $\theta$ and
improved EB
inference by accounting for the uncertainty caused by estimating
$\theta.$

\begin{figure*}[b]

\includegraphics{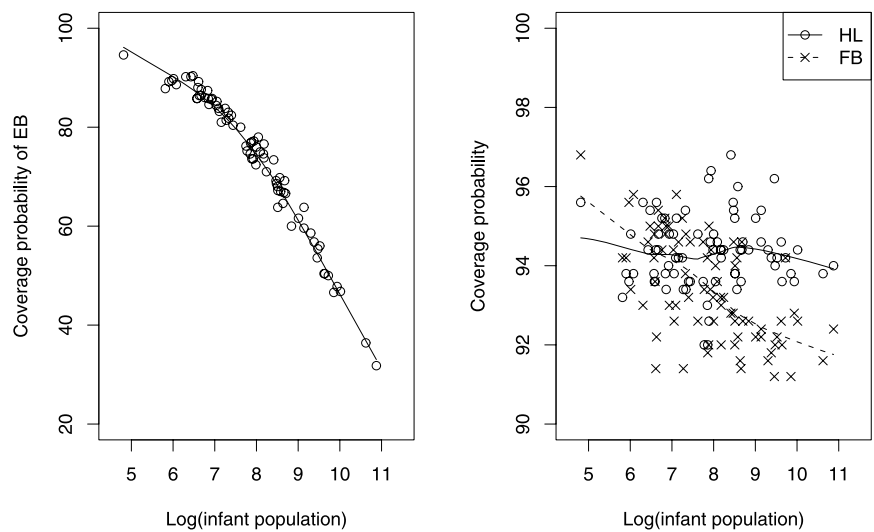}

\caption{ Coverage probabilities of the EB (left) , FB and h-likelihood
(right) methods with respect to population size in the infant mortality
data.}\label{fig4}
\end{figure*}

\subsection{H-likelihood Inferences About $v$}\label{sec4.3}

The example shows that between extended likelihoods $f_{\theta}(y,u)$
and $%
f_{\theta}(y,v)$ the mode of the\break h-likelihood $f_{\theta}(y,v)$
gives a
meaningful estimator for $v$, while that of $f_{\theta}(y,u)$ gives a
meaningless one. Given that extended likelihoods should serve as the basis
for statistical inferences of a general nature, we want to find a particular
scale whose mode gives meaningful inferences about unobservables. Under the
canonical scale the example shows that the mode gives the best
estimator of $%
u$ E$(u|y)$. However, the canonical scale does not exist in general. In
HGLMs Lee and Nelder (\citeyear{2005Lee}) show that maintaining invariance of inference
from extended likelihood for trivial re-expressions of the underlying model
leads to a unique definition of the h-likelihood; we call this the weak
canonical scale in which $v$ appears in the linear predictor.

In Section \ref{sec3.3} we show that APHLs are often similar to marginal
posteriors. Given (marginal) posteriors, a Bayesian would use a
decision-theoretic approach to choose estimators while we use the mode of
the h-likelihood (an extended likelihood on a particular scale) or its
APHLs. Thus the choice of the scale in defining the h-likelihood is
important to guarantee the meaningfulness of the mode estimation. Lee
and Ha
(\citeyear{2008Lee}) show that the standard error estimators from the Hessian matrix (
\ref{eq:hessian}) give the first-order approximation to (\ref{eq:bayes})
with $\pi(\theta)=1$ (Kass and Steffey, \citeyear{1989Kass}) and to the CMSE
(Booth and
Hobert, \citeyear{1998Booth}). Let $w=k(u)$ for some monotone function $k(\cdot)$. Ha and
Lee (\citeyear{2006Lee}) show conditions when the approximation becomes better. One such
condition is that $w|y$ follows the normal distribution. In GLMMs when $v$
is normal we may expect $v|y$ to be approximately normal. If normal the
Laplace approximation is exact; we expect that proposed h-likelihood method
works well. Figure~\ref{sec2} shows how to check the normality of the conditional
distribution by using the APHL.

\subsubsection{Analysis of the BC infant mortality
data}\label{sec4.3.1}

For disease mapping, Leroux et al. (\citeyear{1999Leroux}) and MacNab et al.
(\citeyear{2004MacNab}) consider the conditional autoregressive (CAR) model for the
relative risk $v_{i}$ which satisfies $v\sim N(0,\Sigma) $ where $\Sigma
=\sigma^{2}D^{-1},$ $D=\lambda Q+(1-\lambda)I,$ $\sigma^{2}$ is a
dispersion parameter reflecting the overall heterogeneity of the underlying
risks, and $\lambda$ is a dispersion parameter for the spatial
autocorrelation, $\lambda\in\lbrack0,1].$ The neighborhood matrix
$Q$ has
the $j$th diagonal element equal to the number of neighbors of the
corresponding local region while the off-diagonal elements in each row are
equal to $-1$ if the corresponding regions are neighbors and $0$ otherwise.

The data consist of the number of infant deaths and aggregated mid-year
estimates of the population sizes of infants for $79$ local health areas.
Population size $n_{i}$ varies from $123$ to $52856.$ For these data
Lee et al. (\citeyear{2007Lee}) compare inferences from the h-likelihood with the
full Bayes (FB) analysis. For the FB approach, they set priors $\beta
_{i}\sim N(0,1/0.00001)$ and $\sigma^{-2}\sim\operatorname{gamma}(0.0001,0.0001)$.
Initial values are set as $\sigma^{2}=1$, $\beta_{i}=0$ and
$v_{i}=0$, and
they obtain a posterior sample of 10,000, setting thinning at 10 using
WinBUGS (MacNab et al., \citeyear{2004MacNab}). The coverage probability is
calculated by 95\% Wald confidence intervals based upon asymptotic
normality for the relative risks ($v$) using EB and h-likelihood, and in
the FB method by equal-tail 95\% credible intervals, the interval between
the 2.5th and 97.5th percentiles of the posterior distribution as given by
WinBUGS. For the FB method we use 10,000 iterations after a burn-in of 2000.

Lee et al. (\citeyear{2007Lee}) did a simulation study, assuming $n_{i}$ and
neighborhood structures identical to those in the BC infant mortality; the
data were generated based on (1.1) and (3.1) with $\beta=-4.920,$
$\sigma
^{2}=2$ and $\lambda=0.62$. Using a graph similar to Figure~\ref{fig4}, they showed
that the EB coverage probability decreases dramatically as the population
size $n_{i}$ increases, but that both the h-likelihood and FB methods
improve the EB method substantially by accounting for the uncertainty in
estimating fixed parameters. However, the coverage probability of FB also
decreases as $n_{i}$ increases while the h-likelihood maintains the stated
level of confidence. When $n_{i}$ becomes larger the priors for the
dispersion parameters in the FB may cause problems in frequentist coverage
probability. The h-likelihood procedure maintains the frequentist coverage
probabilities better in this problem. The h-likelihood method is
superior to
Ainsworth and Dean's (\citeyear{2006Ainsworth}) penalized quasi-likelihood (Lee et al.,
\citeyear{2007Lee}) for spatial GLMMs and Ma and Jorgensen's (\citeyear{2007Ma}) orthodox
BLUP\break
method (Lee and Ha, \citeyear{2008Lee}) for nonnormal Tweedie models.


\subsection{Inferences and Model Identifiability}

The joint model for $f_{\theta}(y,v)$ leads to a marginal model
$f_{\theta
}(y)$ for the observed data. We regard $f_{\theta}(y,v)$ as the
fundamental model from which the marginal model can be made. However,
different models for unobservables in $f_{\theta}(y,v)$ can lead to the
same\break marginal model $f_{\theta}(y)$ so that care is necessary in making
inferences about unobservables. Some model assumptions can be checked from
the data while some cannot. This could be an advantage of objective
inference with the likelihood, where uncheckable model assumptions
cannot be
identifiable. In Bayesian analysis, priors can give information on
unidentifiable model assumptions so that it is hard to know whether the
information is coming entirely from the uncheckable priors.

In the modeling of incomplete data we may assume the missing data to be
``missing not at random'' (MNAR) or ``assume random missingness''\break (MAR). Here
assumptions for the missing mechanism cannot be checked by using observed
data [Rubin (\citeyear{2006Rubin})]. Molenberghs et al. (\citeyear{2007Molenberghs}) further show that an
empirical distinction between MAR and MNAR is not possible because each MNAR
model fits to a set of observed data can be reproduced exactly by its
counterpart. Such a pair of models will produce identical estimates for the
observed data but give different estimates for the unobservables (missing
data). Assumptions about unobservables (missing data) are not checkable
without additional information. Unless we have a side-study to determine
whether the observation process depends on what would be observed, all we
have is a model-based assessment. As a referee has pointed out, it will
contain some unverifiable assumptions.

In HGLMs model assumptions for unobservables are often verifiable, that is,
checkable, by using the data because the unobservables are latent variables
for observed data. Consider the one-way random-effect model,
\begin{eqnarray*}
y_{ij}=\beta+v_{i}+e_{ij},
\end{eqnarray*}
where $v_{i}\backsim N(0,\lambda)$ and $e_{ij}\backsim$ $N(0,\phi),$ with
$v_{i}$ and $e_{ij}$ uncorrelated. With more than one observation in each
group the within-group error components $v_{i}$ and $e_{ij}$ are separately
estimable, providing variance-component estimates for the dispersion
parameters. Here model parameters $\phi$ and $\lambda$ connect the
observed data and unobservables. Lee and Nelder\break (\citeyear{2006bLee}) show that if there
are different random-effect models giving the same induced marginal model
for the observed data, then the h-likelihood inferences give equivalent
inferences for equivalent pairs of objects, including unobservables. This
model leads to a marginal model, namely the following
compound-symmetric model:
\begin{eqnarray*}
Y_{i}\sim N(\mathbf{1}\beta,\lambda J_{n_{i}}+\phi I_{n_{i}}).
\end{eqnarray*}
A compound-symmetry model with negative correlation $\lambda<0$ is
perfectly natural in a variety of settings (Nelder, \citeyear{1954Nelder}) which can be
tested by the marginal likelihood (or APHL). Such a model can be
covered by
HGLMs if we allow a negative variance, but then many unanswered questions
arise, such as estimability of random effects, etc.; these require further
research.

Wilk and Kempthorne (\citeyear{1957Wilk}) and Cox (\citeyear{1958Cox}) study the randomization theory
of the Latin square, paying particular attention to the effects on the
interpretation of the conventional analysis of variance (ANOVA) of the
absence of unit-treatment additivity, a point first raised by Neyman (\citeyear{1935Neyman}).
Consider a model for the Latin-square design,
%
\begin{eqnarray}\label{eq:latin1}
y_{ij(k)}&=&\mu+r_{i}+c_{j}+\tau_{k}\nonumber
\\[-8pt]\\[-8pt]
&&{}+(rc)_{ij}+(rt)_{ik}+(ct)_{jk}+e_{ij(k)}.\nonumber
\end{eqnarray}
Suppose that the main effects are regarded as fixed. When the
interactions $%
(rc)_{ij},(rt)_{ik}, (ct)_{jk}$ are fixed a test for the main
effect is
irrelevant because it makes no sense to postulate that either of the two
main effects is null when their interaction is not assumed zero
(Nelder, \citeyear{1994Nelder}). However, if the interactions are regarded as random the associated
main effects can tested without any difficulty from the ANOVA table.
Permutation from~a~fi\-nite population is a way of generating distributions
for random effects. Wilk and Kempthorne (\citeyear{1957Wilk}) put constraints $%
\sum_{i}(rc)_{ij}=\sum_{j}(rc)_{ij}=\cdots=\sum_{k}(ct)_{jk}=0.$
Nelder
(\citeyear{1994Nelder}) points out that such constraints make no sense either with fixed or
random effects. With fixed effects the choice of constraints to give the
least-square equations a solution is essentially arbitrary. However, with
random effects symmetric constraints on estimates of the parameters of the
form $\widehat{\sum_{i}(rc)_{ij}}=\sum_{j}\widehat{(rc)_{ij}}=\cdots=
\sum_{k}%
\widehat{(ct)_{jk}}=0$ arise naturally (Lee and Nelder, \citeyear{1996Lee}, \citeyear{2005Lee}).
However, here only fractions of combinations are used to make the combined
error component $v_{ij(k)}=(rc)_{ij}+(rt)_{ik}+(ct)_{jk}+e_{ij(k)}$ to form
a sum of independent errors. Thus model (\ref{eq:latin1}) gives an
identical marginal model to the conventional model for Latin squares with
main effects only
%
\begin{equation}\label{eq:latin2}
y_{ij(k)}=\mu+r_{i}+c_{j}+\tau_{k}+e_{ij(k)}^{\ast}.
\end{equation}
From Lee and Nelder (\citeyear{2006bLee}) the two models lead to identical inferences
about both fixed parameters and random effects, giving $\hat{e}%
_{ij(k)}^{\ast}=\hat{v}_{ij(k)}.$ Thus in (\ref{eq:latin1}) individual
error components cannot be separated by the observed data. If a method can
identify individual components then it must be based upon uncheckable model
assumptions such as priors. Consider the following model:
%
\begin{equation}
y_{ij(k)}=\mu+r_{i}+c_{j}+\tau_{ij(k)}+e_{ij(k)},
\end{equation}
where $\tau_{ij(k)}=\tau_{k}+(rt)_{ik}+(ct)_{jk}$ and $(rt)_{ik}$ and
$%
(ct)_{jk}$ are random with zero means. This model assumes unit-treatment
interaction and can be interpreted to have the average treatment effects
such that
\begin{eqnarray*}
\mbox{E}\bigl(\tau_{ij(k)}\bigr)=\tau_{k}.
\end{eqnarray*}
Then we can test that the average treatment effects are the same (Lee and
Nelder, \citeyear{2002Lee}). Thus with unobservables there are different methods of
interpretation: we may consider $(rt)_{ik}$ and $(ct)_{jk}$ to be either
error components or random treatment-unit interactions. These give
equivalent inferences for equivalent quantities.

\subsection{Discussion}

There have been many alleged examples similar to that of Bayarri et
al. (\citeyear{1988Bayarri}) and Little and Rubin [(\citeyear{2002Little}), Chapter 6.3], purporting to show
that an extension of the Fisher likelihood to three kinds of objects is not
possible. Lee and Nelder (\citeyear{2005Lee}) refute those of Bayarri et al. and
Yun et al. (\citeyear{2007Yun}) those of Little and Rubin. These complaints are,
we believe, resolved by the h-likelihood framework. Zhao et al.
(2006) claim that the Bayesian analysis is computationally simpler for
obtaining variance estimators for the random-effect estimates compared with
its frequentist counterpart; however with the extended likelihood framework
this may not be so, at least in the analysis of the disease-mapping
areas in
Section \ref{sec4.3.1}.

The h-likelihood (\ref{eq:h-likelihood}) gives a new definition of conjugate
families (Lee and Nelder, \citeyear{2001aLee}), showing that the likelihood for a conjugate
family for $\log f_{\theta}(v)$ takes the form of a GLM. It is the sum of
component likelihoods, $\log f_{\theta}(v)$ and $\log f_{\theta}(y|v),$
both representable as GLM likelihoods. This means that an extended
class of
models can be decomposed into component GLMs (Lee and Nelder, \citeyear{2001aLee},
\citeyear{2006aLee})
and that these extended models can be fitted as an interconnected set of
component GLMs. This greatly facilitates the development of model-checking
techniques for the whole class (Lee and Nelder, \citeyear{2001aLee}). A single algorithm,
iterative weighted least squares, can be used throughout all this extended
class of models and requires neither prior distributions of parameters nor
multi-dimensional quadrature. The h-likelihood plays a key role in the
\mbox{synthesis} of the computational algorithms needed for this extended
class of
models.

This formulation means that a great variety of models can be fitted by a
single algorithm and compared using extensions of standard GLM procedures.
Thus we can change the link function, allow various types of term in the
linear predictor and use model-selection methods for adding or deleting
terms. Furthermore, various model assumptions can be checked by applying
GLM model-checking procedures to the appropriate component GLMs. This
establishes, we believe, algorithmic \textit{wiseness} in the sense of Efron
(\citeyear{2003Efron}).

\section{Conclusion}\label{sec5}

We have shown that a broad class of new models with wide applications
can be
generated by the probabilistic modeling of unobservables. There has
been an
attempt using the GEE method to make inferences from general non-normal
multivariate models without modeling unobservables. It pre-empts model
selection by claiming to make inferences about population averages or
marginal means. We do not disagree with the need to make marginal
predictions after choosing a model but believe that such a need does not
require, and indeed should not use, prediction methods at the
model-selection stage. We dislike the pre-emption of the model selection
stage by a particular prediction method. Furthermore, these population,
marginal and subject-specific averages are parameterizations in the
probabilistic model.\break When a prediction method lacks a probabilistic model
basis it is not possible to connect these parameters and compare them.

We do not object to the use of Fisher's likelihood for inferences about
fixed parameters. The Fisher likelihood framework has advantages such as
generality of application, statistical and computational efficiency,
etc., and
we agree with its use. However, it cannot deal with inferences from models
having unobservables because there is always a problem of inference about
those unobservables. H-likelihood gives a powerful and practical framework
for statistical inference of general model class with unobservables,
maintaining the advantages of the original likelihood framework for fixed
parameters. We believe that more new classes of models will be
developed and
that the h-likelihood will become widely used for inference from them.

The h-likelihood uses the mode and its curvature for inferences about
unobservables. Thus, in defining the h-likelihood the scale of unobservables
must be carefully chosen to make a valid inferences. The (weak) canonical
scale in HGLMs leads to an invariance of a certain extended likelihood.
However, in general the validity of such a scale has not been established.
The conditional normality in Section~\ref{sec4.3} would be a promising condition to
determine the scale, which can be checked by plotting the APHL. Further
studies are required on the scale in defining the h-likelihood under general
situations beyond DHGLMs. For fixed parameter estimation we use the marginal
likelihood. But it is often hard to compute, so that we have proposed using
the Laplace approximation. However, this approximation gives nonnegligible
biases in binary data. We have found that the second-order
approximation is
effective in eliminating such biases. However, it becomes very hard to
implement as the number of random components increases. So it would be of
interest to find an approximation which can be implemented under general
situations.

\section*{Acknowledgments}
The authors thank Professors Jan Bj\o rnstad, Martin Crowder, Harry Joe,
Jaeyong Lee, Yudi Pawitan and Roger Payne for their helpful comments. This
work was supported by Brain Korea 21.

\end{document}